\documentclass[%
 reprint,
superscriptaddress,
 amsmath,amssymb,
 aps,
]{revtex4-2}

\usepackage{graphicx}
\usepackage{dcolumn}
\usepackage{bm}
\usepackage[mathlines]{lineno}
\usepackage{multirow}
\usepackage{tabularray}
\usepackage{comment}
\usepackage{xcolor}

\begin{document}

\preprint{APS/123-QED}

\title{Search for proton decay into a single charged antilepton and a massless invisible particle using the full pure water data set of Super-Kamiokande}

\newcommand{\AFFicrr}{\affiliation{Kamioka Observatory, Institute for Cosmic Ray Research, University of Tokyo, Kamioka, Gifu 506-1205, Japan}}
\newcommand{\AFFkashiwa}{\affiliation{Research Center for Cosmic Neutrinos, Institute for Cosmic Ray Research, University of Tokyo, Kashiwa, Chiba 277-8582, Japan}}
\newcommand{\AFFipmu}{\affiliation{Kavli Institute for the Physics and
Mathematics of the Universe (WPI), The University of Tokyo Institutes for Advanced Study,
University of Tokyo, Kashiwa, Chiba 277-8583, Japan }}
\newcommand{\AFFmad}{\affiliation{Department of Theoretical Physics, University Autonoma Madrid, 28049 Madrid, Spain}}
\newcommand{\AFFubc}{\affiliation{Department of Physics and Astronomy, University of British Columbia, Vancouver, BC, V6T1Z4, Canada}}
\newcommand{\AFFbu}{\affiliation{Department of Physics, Boston University, Boston, MA 02215, USA}}
\newcommand{\AFFuci}{\affiliation{Department of Physics and Astronomy, University of California, Irvine, Irvine, CA 92697-4575, USA }}
\newcommand{\AFFcsu}{\affiliation{Department of Physics, California State University, Dominguez Hills, Carson, CA 90747, USA}}
\newcommand{\AFFcnm}{\affiliation{Institute for Universe and Elementary Particles, Chonnam National University, Gwangju 61186, Korea}}
\newcommand{\AFFduke}{\affiliation{Department of Physics, Duke University, Durham NC 27708, USA}}
\newcommand{\AFFgifu}{\affiliation{Department of Physics, Gifu University, Gifu, Gifu 501-1193, Japan}}
\newcommand{\AFFgist}{\affiliation{GIST College, Gwangju Institute of Science and Technology, Gwangju 500-712, Korea}}
\newcommand{\AFFuh}{\affiliation{Department of Physics and Astronomy, University of Hawaii, Honolulu, HI 96822, USA}}
\newcommand{\AFFicl}{\affiliation{Department of Physics, Imperial College London , London, SW7 2AZ, United Kingdom }}
\newcommand{\AFFkek}{\affiliation{High Energy Accelerator Research Organization (KEK), Tsukuba, Ibaraki 305-0801, Japan }}
\newcommand{\AFFkobe}{\affiliation{Department of Physics, Kobe University, Kobe, Hyogo 657-8501, Japan}}
\newcommand{\AFFkyoto}{\affiliation{Department of Physics, Kyoto University, Kyoto, Kyoto 606-8502, Japan}}
\newcommand{\AFFliv}{\affiliation{Department of Physics, University of Liverpool, Liverpool, L69 7ZE, United Kingdom}}
\newcommand{\AFFmiyagi}{\affiliation{Department of Physics, Miyagi University of Education, Sendai, Miyagi 980-0845, Japan}}
\newcommand{\AFFnagoya}{\affiliation{Institute for Space-Earth Environmental Research, Nagoya University, Nagoya, Aichi 464-8602, Japan}}
\newcommand{\AFFkmi}{\affiliation{Kobayashi-Maskawa Institute for the Origin of Particles and the Universe, Nagoya University, Nagoya, Aichi 464-8602, Japan}}
\newcommand{\AFFpol}{\affiliation{National Centre For Nuclear Research, 02-093 Warsaw, Poland}}
\newcommand{\AFFsuny}{\affiliation{Department of Physics and Astronomy, State University of New York at Stony Brook, NY 11794-3800, USA}}
\newcommand{\AFFokayama}{\affiliation{Department of Physics, Okayama University, Okayama, Okayama 700-8530, Japan }}
\newcommand{\AFFosaka}{\affiliation{Department of Physics, Osaka University, Toyonaka, Osaka 560-0043, Japan}}
\newcommand{\AFFox}{\affiliation{Department of Physics, Oxford University, Oxford, OX1 3PU, United Kingdom}}
\newcommand{\AFFqmul}{\affiliation{School of Physics and Astronomy, Queen Mary University of London, London, E1 4NS, United Kingdom}}
\newcommand{\AFFregina}{\affiliation{Department of Physics, University of Regina, 3737 Wascana Parkway, Regina, SK, S4SOA2, Canada}}
\newcommand{\AFFseoul}{\affiliation{Department of Physics and Astronomy, Seoul National University, Seoul 151-742, Korea}}
\newcommand{\AFFsheff}{\affiliation{School of Mathematical and Physical Sciences, University of Sheffield, S3 7RH, Sheffield, United Kingdom}}
\newcommand{\AFFshizuokasc}{\affiliation{Department of Informatics in
Social Welfare, Shizuoka University of Welfare, Yaizu, Shizuoka, 425-8611, Japan}}
\newcommand{\AFFstfc}{\affiliation{STFC, Rutherford Appleton Laboratory, Harwell Oxford, and Daresbury Laboratory, Warrington, OX11 0QX, United Kingdom}}
\newcommand{\AFFskk}{\affiliation{Department of Physics, Sungkyunkwan University, Suwon 440-746, Korea}}
\newcommand{\AFFtodai}{\affiliation{Department of Physics, University of Tokyo, Bunkyo, Tokyo 113-0033, Japan }}
\newcommand{\AFFtit}{\affiliation{Department of Physics, Institute of Science Tokyo, Meguro, Tokyo 152-8551, Japan }}
\newcommand{\AFFtus}{\affiliation{Department of Physics and Astronomy, Faculty of Science and Technology, Tokyo University of Science, Noda, Chiba 278-8510, Japan }}
\newcommand{\AFFtriumf}{\affiliation{TRIUMF, 4004 Wesbrook Mall, Vancouver, BC, V6T2A3, Canada }}
\newcommand{\AFFtokai}{\affiliation{Department of Physics, Tokai University, Hiratsuka, Kanagawa 259-1292, Japan}}
\newcommand{\AFFtsinghua}{\affiliation{Department of Engineering Physics, Tsinghua University, Beijing, 100084, China}}
\newcommand{\AFFynu}{\affiliation{Department of Physics, Yokohama National University, Yokohama, Kanagawa, 240-8501, Japan}}
\newcommand{\AFFllr}{\affiliation{Ecole Polytechnique, IN2P3-CNRS, Laboratoire Leprince-Ringuet, F-91120 Palaiseau, France }}
\newcommand{\AFFbari}{\affiliation{ Dipartimento Interuniversitario di Fisica, INFN Sezione di Bari and Universit\`a e Politecnico di Bari, I-70125, Bari, Italy}}
\newcommand{\AFFnapoli}{\affiliation{Dipartimento di Fisica, INFN Sezione di Napoli and Universit\`a di Napoli, I-80126, Napoli, Italy}}
\newcommand{\AFFroma}{\affiliation{INFN Sezione di Roma and Universit\`a di Roma ``La Sapienza'', I-00185, Roma, Italy}}
\newcommand{\AFFpadova}{\affiliation{Dipartimento di Fisica, INFN Sezione di Padova and Universit\`a di Padova, I-35131, Padova, Italy}}
\newcommand{\AFFkeio}{\affiliation{Department of Physics, Keio University, Yokohama, Kanagawa, 223-8522, Japan}}
\newcommand{\AFFwinnipeg}{\affiliation{Department of Physics, University of Winnipeg, MB R3J 3L8, Canada }}
\newcommand{\AFFkcl}{\affiliation{Department of Physics, King's College London, London, WC2R 2LS, UK }}
\newcommand{\AFFwarwick}{\affiliation{Department of Physics, University of Warwick, Coventry, CV4 7AL, UK }}
\newcommand{\AFFral}{\affiliation{Rutherford Appleton Laboratory, Harwell, Oxford, OX11 0QX, UK }}
\newcommand{\AFFwu}{\affiliation{Faculty of Physics, University of Warsaw, Warsaw, 02-093, Poland }}
\newcommand{\AFFbcit}{\affiliation{Department of Physics, British Columbia Institute of Technology, Burnaby, BC, V5G 3H2, Canada }}
\newcommand{\AFFtohoku}{\affiliation{Department of Physics, Faculty of Science, Tohoku University, Sendai, Miyagi, 980-8578, Japan }}
\newcommand{\AFFicise}{\affiliation{Institute For Interdisciplinary Research in Science and Education, ICISE, Quy Nhon, 55121, Vietnam }}
\newcommand{\AFFilance}{\affiliation{ILANCE, CNRS - University of Tokyo International Research Laboratory, Kashiwa, Chiba 277-8582, Japan}}
\newcommand{\AFFibs}{\affiliation{Center for Underground Physics, Institute for Basic Science (IBS), Daejeon, 34126, Korea}}
\newcommand{\AFFglasgow}{\affiliation{School of Physics and Astronomy, University of Glasgow, Glasgow, Scotland, G12 8QQ, United Kingdom}}
\newcommand{\AFFoecu}{\affiliation{Media Communication Center, Osaka Electro-Communication University, Neyagawa, Osaka, 572-8530, Japan}}
\newcommand{\AFFminn}{\affiliation{School of Physics and Astronomy, University of Minnesota, Minneapolis, MN  55455, USA}}
\newcommand{\AFFsilesia}{\affiliation{August Che\l{}kowski Institute of Physics, University of Silesia in Katowice, 75 Pu\l{}ku Piechoty 1, 41-500 Chorz\'{o}w, Poland}}
\newcommand{\AFFtoyama}{\affiliation{Faculty of Science, University of Toyama, Toyama City, Toyama 930-8555, Japan}}
\newcommand{\AFFbmcc}{\affiliation{Science Department, Borough of Manhattan Community College / City University of New York, New York, New York, 1007, USA.}}
\newcommand{\AFFnumazu}{\affiliation{National Institute of Technology, Numazu College, Numazu, Shizuoka  410-8501, Japan}}
\newcommand{\AFFniihama}{\affiliation{National Institute of Technology, Niihama College, Niihama, Ehime  792-8580, Japan}}
\newcommand{\AFFucas}{\affiliation{School of Physical Sciences, University of Chinese Academy of Sciences, Beijing 101408, China}}

\AFFicrr
\AFFkashiwa
\AFFmad
\AFFbmcc
\AFFbu
\AFFbcit
\AFFuci
\AFFcsu
\AFFucas
\AFFcnm
\AFFduke
\AFFllr
\AFFgifu
\AFFgist
\AFFglasgow
\AFFuh
\AFFibs
\AFFicise
\AFFicl
\AFFbari
\AFFnapoli
\AFFpadova
\AFFroma
\AFFilance
\AFFkeio
\AFFkek
\AFFkcl
\AFFkobe
\AFFkyoto
\AFFliv
\AFFminn
\AFFmiyagi
\AFFnagoya
\AFFkmi
\AFFpol
\AFFniihama
\AFFnumazu
\AFFsuny
\AFFokayama
\AFFoecu
\AFFox
\AFFral
\AFFseoul
\AFFsheff
\AFFshizuokasc
\AFFsilesia
\AFFstfc
\AFFskk
\AFFtohoku
\AFFtodai
\AFFipmu
\AFFtit
\AFFtus
\AFFtoyama
\AFFtriumf
\AFFtsinghua
\AFFwu
\AFFwarwick
\AFFwinnipeg
\AFFynu

\author{Y.~M.~Liu}
\AFFkeio
\author{K.~Terada}
\AFFtit
\author{K.~Abe}
\AFFicrr
\AFFipmu
\author{Y.~Asaoka}
\AFFicrr
\AFFipmu
\author{M.~Harada}
\AFFicrr
\author{Y.~Hayato}
\AFFicrr
\AFFipmu
\author{K.~Hiraide}
\AFFicrr
\AFFipmu
\author{T.~H.~Hung}
\AFFicrr
\author{K.~Ieki}
\author{M.~Ikeda}
\AFFicrr
\AFFipmu
% ...


\collaboration{The Super-Kamiokande Collaboration}
\noaffiliation

\date{\today}


\begin{abstract}
A search for proton decay via $p\rightarrow l^{+}+X$, where $l^{+}$ is a positively charged lepton and $X$ is an invisible, massless, neutral particle, was performed using a 401~kton$\cdot$years exposure representing the entire pure water phase of Super-Kamiokande. 
No significant indication of a proton decay was observed beyond the expected atmospheric neutrino background. 
Lower limits on the partial lifetime of the proton were set to at $1.72\times10^{33}$ years for $p\rightarrow e^{+}+X$ and $0.61\times10^{33}$ years for $p\rightarrow \mu^{+}+X$ at the $90\%$ confidence level. 
These results improve on previous limits by factors of 2 and 1.5, respectively. 
\end{abstract}

\maketitle


\section{\label{sec:level1}Introduction}
The Standard Model (SM) is currently the best description of elementary particles and their interactions, explaining most experimental observations with remarkable precision. 
Grand Unified Theories (GUTs)~\cite{PhysRevD.8.1240, PhysRevLett.31.661, PhysRevLett.32.438, FRITZSCH1975193}, on the other hand, are proposed extensions of the SM that are intended to provide natural explanations to questions that the SM does not fully address, such as the nature of charge quantization and the current observed matter--antimatter asymmetry of the universe. 
Prototypical GUTs unify the strong, weak and electromagnetic interactions at energy scales on the order of $10^{15}$--$10^{16}$ GeV, energies that are unreachable by accelerators. 
However, large underground detectors can search for rare processes predicted by GUTs, such as proton decay, to probe these models and energy scales. 
The observation of an unstable proton is a predicted signature of these models that would constitute strong evidence for new physics via its violation of baryon number.

The simplest unification models are represented by the minimal SU(5)~\cite{LANGACKER1981185} and the minimal supersymmetric (SUSY) SU(5)~\cite{HISANO199346}. 
They predict proton decays dominantly via $p\rightarrow e^{+}+\pi^{0}$~\cite{PhysRevD.59.052004, HIRATA1989308, PhysRevLett.81.3319} and $p\rightarrow \overline{\nu}+K^{+}$~\cite{PhysRevD.90.072005}, respectively. 
However, observations by Super-Kamiokande (SK) as well as earlier experiments place lower limits on the proton lifetime that rule out these predictions.
More elaborate models have been proposed, such as those based on SO(10)~\cite{PhysRevLett.70.2845, GOH2004105} (with or without SUSY) make predictions that exceed current experimental bounds and can additionally explain the pattern of fermion masses and mixings~\cite{PhysRevD.64.053015, PhysRevD.65.079904}.
Models with additional light invisible particles, such as sterile neutrinos, axion-like particles, or dark scalars, can predict proton decays where only part of the energy is distributed among SM particles~\cite{PhysRevD.110.L031701}.  
Searches for $p\rightarrow$ ($l^{+}+$missing energy)~\cite{Helo_2018} decays therefore, offer an alternative window onto unification and baryon number violation with the potential to additionally probe such new light particles.

In this study, we report latest results for a search for proton decays via the two-body process, $p\rightarrow l^{+}+X$ in the SK detector. 
We focus on $p\rightarrow e^{+}+X$ and $p\rightarrow \mu^{+}+X$, where $X$ is an unknown invisible massless neutral particle, updating our previous search~\cite{PhysRevLett.115.121803}.
That work set lower limits on the proton lifetime as $7.9\times10^{32}$ years for $p\rightarrow e^{+}+X$ and $4.1\times10^{32}$ years for $p\rightarrow \mu^{+}+X$ at the $90\%$ confidence level (C.L.), respectively.
In this study, the experimental exposure is extended from 273.4 kton$\cdot$years to 401 kton$\cdot$years, including all pure-water data taken at SK since 1996.

This paper is structured as follows: The Super-Kamiokande detector is described in Section II and the simulation methods employed in this search are outlined in Section III. 
Event reconstruction methods are introduced in Section IV. 
Selection criteria, analysis results, and systematic uncertainties are presented in Sections V, VI, and VII. 
Section VIII presents a summary of this work.

\section{The super-Kamiokande experiment}
Super-Kamiokande is a large water Cherenkov detector located in the Kamioka mine in Hida City, Gifu Prefecture, Japan~\cite{FUKUDA2003418}. 
It is located 1,000 m below the peak of Mt. Ikenoyama, which is a depth of approximately a 2,700 m water equivalent, to reduce the cosmic ray muon background. 
The detector is housed in a stainless steel tank, 39.3 m in diameter and 41.4 m in height. 
It is filled with 50 kton of ultra pure water. 
The Inner Detector (ID) volume, 33.8 m in diameter and 36.2 m in height, is covered with more than 11,000 inward-facing 20-inch photomultiplier tubes (PMTs) and contains 32 kton of water. 
The 2 m-thick Outer Detector (OD) is built around the outside of the ID and is covered with 1885 outward-facing 8-inch PMTs.
Reflective Tyvek lines the OD surfaces to increase light collection. 
The OD serves two functions: vetoing the cosmic ray muon background and shielding against gamma rays from surrounding rock.
The 55~cm region between the ID and OD is insensitive. 

Pure water data collection phases at SK are classified into five distinct periods, SK-I to SK-V. 
Each phase has similar operating conditions, with a few notable exceptions. 
SK-I operation began in 1996 and ended in 2001 for maintenance purposes. 
An accident during the maintenance process led to the loss of approximately half of the PMTs in the detector. 
Fiber reinforced plastic and acrylic PMT cases were thereafter used to prevent further accidents, starting with SK-II.
The remaining PMTs were rearranged and SK-II was operated from 2002 to 2005 with the ID photocoverage reduced from 40 to 19$\%$. 
SK-III ran from 2006 to 2008, with new PMTs installed to recover those lost in the accident. 
The readout electronics and data acquisition system were upgraded at the start of the SK-IV phase~\cite{NISHINO2009710}. 
The upgraded electronics extended the window of hit times recorded following neutrino-like events, which improved the efficiency of detecting Michel electrons~\cite{5446533}. 
SK-IV operated from 2008 to 2018. 
In 2018, the tank walls were sealed to prevent leaks and around 140 ID PMTs were replaced. 
This marked the start of the SK-V phase, which collected data from 2019 to 2020. 
Gadolinium was added to the detector water in July 2020~\cite{ABE2022166248} in order to improve neutron detection.
Since that time SK has collected data as SK-Gd.
The present study does not include data from the SK-Gd phase.

In this analysis, detected particles are considered Fully Contained (FC) in the ID, with the reconstructed vertex inside the Fiducial Volume (FV). 
The FV is defined as the detector volume located least 2 m away from the inner surface of the ID and corresponds to 22.5 kton of pure water. 
This analysis builds upon the search performed in Ref.~\cite{PhysRevLett.115.121803}, adopting the Monte Carlo (MC) simulations, the fit methodology, and the processed SK-I to SK–III data developed in that work. 
Herein, SK-IV and -V are newly analyzed to extend the exposure to the entire pure water phase of the experiment. 
With this extension, the SK-IV exposure is expanded from 100.56 kton$\cdot$years to 199.86 kton$\cdot$years, and the 28.40~kton$\cdot$year SK-V data is analyzed for the first time. 
Whereas Ref.~\cite{PhysRevLett.115.121803} considered only systematics with an expected impact on any analysis bin of 0.05 or larger, the present work utilizes all systematics to provide a more comprehensive uncertainty model and re-estimates error sizes for the SK-I to SK-III data.

\section{Simulation}
Monte Carlo detector simulations for the proton decay signal and atmospheric neutrino backgrounds are utilized to estimate the detection performance and for comparison between the simulated expectation and experimental observation. 
Owing to variations in the detector configurations (SK-I to SK–V), MC samples are generated separately in each SK period.

Since the SK target is water, there are two unbound (``free'') protons in the hydrogen atoms of each molecule and eight protons bound in the oxygen atom.
Each proton is assumed to decay with equal probability. 
Protons in hydrogen are simulated assuming an initial mass of 938.27 MeV/$c^{2}$ and a momentum 0 MeV/$c$ in the lab frame.
Each daughter particle ($l^{+}$ and $X$) decaying from a free proton takes half of the rest mass energy of the proton, and 
their directions are back-to-back. 
A uniform phase space is assumed for the decay kinematics of the outgoing charged lepton with no additional correlations being taken into account. 

\begin{figure}[!htbp]
    \centering
    \includegraphics[width=.4\textwidth]{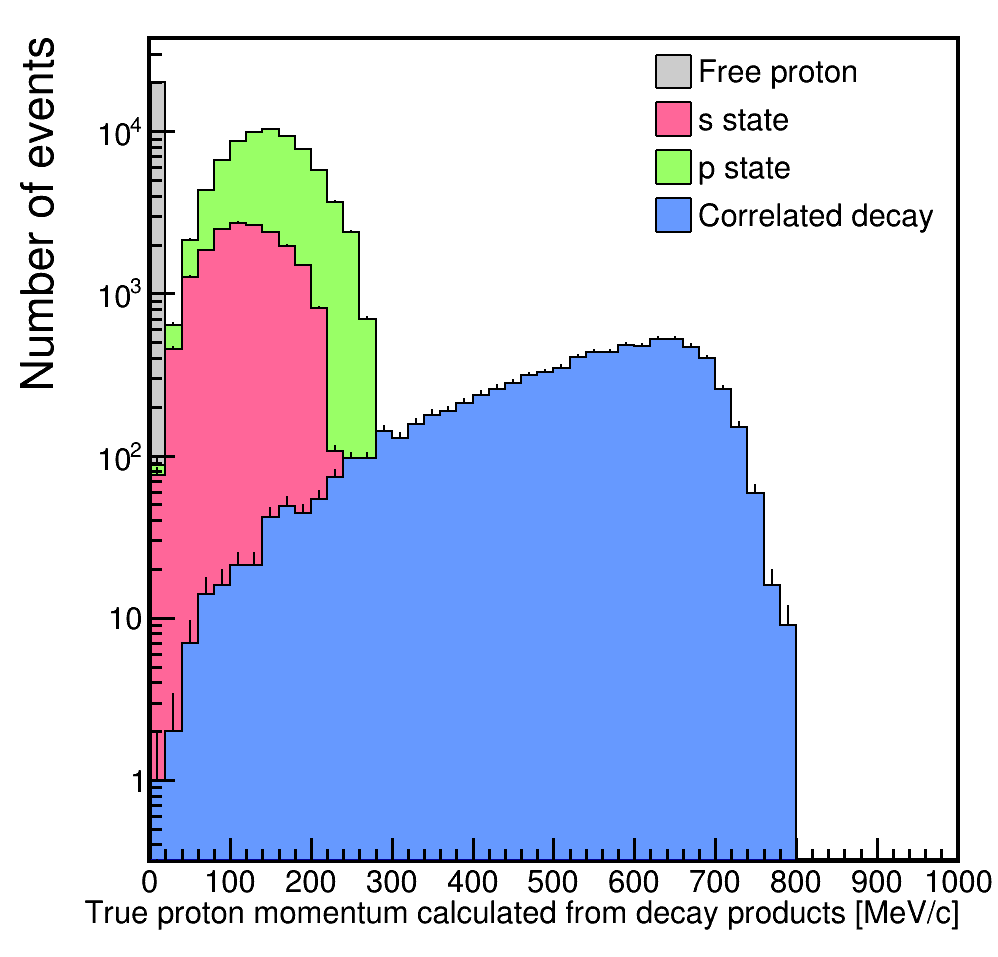}
    \includegraphics[width=.4\textwidth]{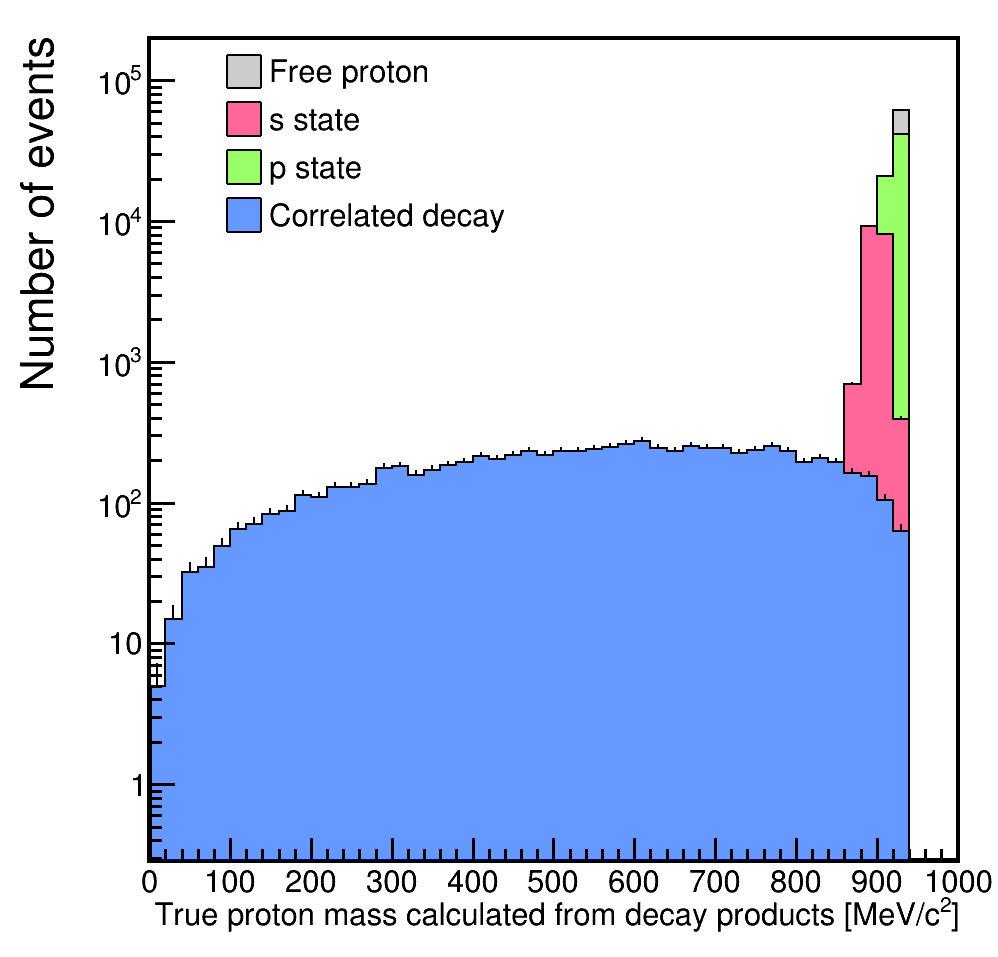}
    \caption{True proton momentum distribution (top) and the mass distribution (bottom), with the components of the free protons, bound protons ($s$ and $p$ states), and correlated decay. Note that the colored histograms are stacked, and the true momentum and mass are calculated from the decay products, $e^{+}$ and $X$.}
    \label{fig:proton}
\end{figure}

Protons in oxygen, on the other hand, carry some initial momentum determined by their Fermi motion. 
The Fermi momentum assumed in the simulation is based on the spectral function derived from the measurement of electron-$^{12}$C scattering~\cite{nakamura_hiramatsu_kamae_muramatsu_izutsu_watase_1976}. 
For a bound proton decay, the initial state is assigned to either an $s$ or a $p$ state in a 1:3 ratio based on the nuclear shell model~\cite{mayer1955elementary}. 
Its effective mass is calculated by subtracting the binding energy from the proton rest mass.
Binding energies are taken at random from assuming a Gaussian distribution with mean of 39.0~MeV and standard deviation of 10.2~MeV for the $s$ state and 15.5~MeV and 3.8~MeV for the $p$ state~\cite{shoda1972nuclear}.
Owing to the overlapping wave function of bound protons in the nuclear environment, an additional ``spectator'' nucleon is involved in the decay and thus modifies the decay system. 
Such ``correlated decays'' have their kinematics broadened by the presence of the ``spectator'' nucleon. 
The probability of this correlated decay effect is predicted to be 10$\%$~\cite{YAMAZAKI19991}. 
Figure~\ref{fig:proton} shows the true proton momentum and mass distributions for the $e^{+}$ and $X$ decay mode, 
showing the free proton, bound proton ($s$ and $p$ states), and correlated decay components.

Atmospheric neutrinos represent the largest background to proton decay searchers.
The atmospheric neutrino flux used in the MC simulation is calculated using the model by Honda $et$ $al.$~\cite{PhysRevD.75.043006, PhysRevD.83.123001}. 
Neutrino interactions with hydrogen or oxygen nuclei are simulated using NEUT~\cite{hayato2021}, which is based on the Fermi gas model~\cite{PhysRevC.83.045501}. 
Atmospheric neutrino samples equivalent to a 500-year exposure of each SK period are used in this analysis.

The SK detector simulation includes Cherenkov radiation processes, the propagation of particles and Cherenkov photons in water, and the responses of the PMTs and their electronic. 
The detector simulation is based on the GEANT-3 package~\cite{Brun:1082634}. 
Calibrations are performed using several control samples as detailed in Ref.~\cite{FUKUDA2003418, ABE2014253}.

\section{Event reconstruction}
Reconstruction processes are applied for events that remain after the FC reduction processes, which selects events with little OD PMT activity and more than 30~MeV of visible energy in the ID while discarding entering backgrounds and non-physics noise events.
An event's kinematic parameters, such as the interaction vertex, number of Cherenkov rings, the particle ID (PID) of each ring, their reconstructed momenta and the number of delayed Michel electrons, are determined by using the PMT hit positions, charge, and time information. 
This analysis uses the APFit~\cite{SHIOZAWA1999240, doi:10.1142/9789819801107_0008} algorithm for processing both MC and data events.

The reconstruction begins by determining the vertex, assuming that the Cherenkov light arriving at the ID PMTs originates from the same point.
The direction and edge of the most likely Cherenkov ring are then estimated based on the angular distribution of the observed charge. 
Cherenkov rings are searched for using a pattern recognition algorithm based on the Hough transform~\cite{davies2004machine} using the charge distribution for each hit PMT. 
Reconstructed Cherenkov rings can be categorized as either showering ($e$-like for $e^{\pm}$, $\gamma$) or non-showering ($\mu$-like for $\mu^{\pm}$, $\pi^{\pm}$) rings. 
Cherenkov rings generated by $e$-like events produce fuzzy rings owing to electromagnetic scattering and showering (Bremsstrahlung $e^{\pm}\rightarrow e^{\pm}+\gamma$). Conversely, Cherenkov rings generated by $\mu$-like events have much clearer edges. 
Ring types can therefore be identified on the basis of a likelihood algorithm based on the pattern of the observed PMT hits and expected charge distribution. 
For single-Cherenkov-ring events, the Cherenkov opening angle is also considered for particle identification. 
The vertex position is refined using this fitted ring information to improve the position resolution.

Each Cherenkov ring's momentum is calculated using the integrated photoelectrons observed within a 70$^{\circ}$ cone around the reconstructed direction of the ring.
The conversion from photoelectrons to momentum is based on simulations of single electrons and muons sampled over many input momenta. 
PID is considered when reconstructing the ring momentum.
Michel electrons are tagged as PMT hit clusters separated in time after the primary event's PMT hits. 
More details on the reconstruction method can be found in Ref.~\cite{doi:10.1142/9789819801107_0008, SHIOZAWA1999240}.

\section{Event selection}
The event selection performance is studied using signal and background MC simulation events. 
Here 100,000 signal events for each decay mode are generated uniformly in the entire ID volume. 
These generated signal and the atmospheric neutrino background events are first selected by the FCFV pre-selection cuts described above~\cite{PhysRevD.95.012004,PhysRevD.90.072005,PhysRevD.96.012003}.

The following selection criteria are then applied to events passing the FCFV selection to extract proton decay signals separately from atmospheric neutrino background events. 
The selection criteria are the same as those in the previous analysis~\cite{PhysRevLett.115.121803}.

\vskip\baselineskip
(i) The visible energy of events is in the sub-GeV region ($<$1330 MeV).

(ii) The number of reconstructed Cherenkov rings must be only one.

(iii) The ring must be an electron-like ring for the $p\rightarrow e^{+}+X$ mode and a muon-like ring for the $p\rightarrow \mu^{+}+X$ mode.

(iv) An event with zero Michel electrons is required for the $p\rightarrow e^{+}+X$ mode, and one Michel electron is required for the $p\rightarrow \mu^{+}+X$ mode.

(v) The reconstructed momentum of the $e$-like ring events should be in the range $100-1000$ MeV/c, and in the range $200-1000$ MeV/c for the muon-like ring events.
\vskip\baselineskip

\begin{figure}[!htbp]
    \centering
    \includegraphics[width=.4\textwidth]{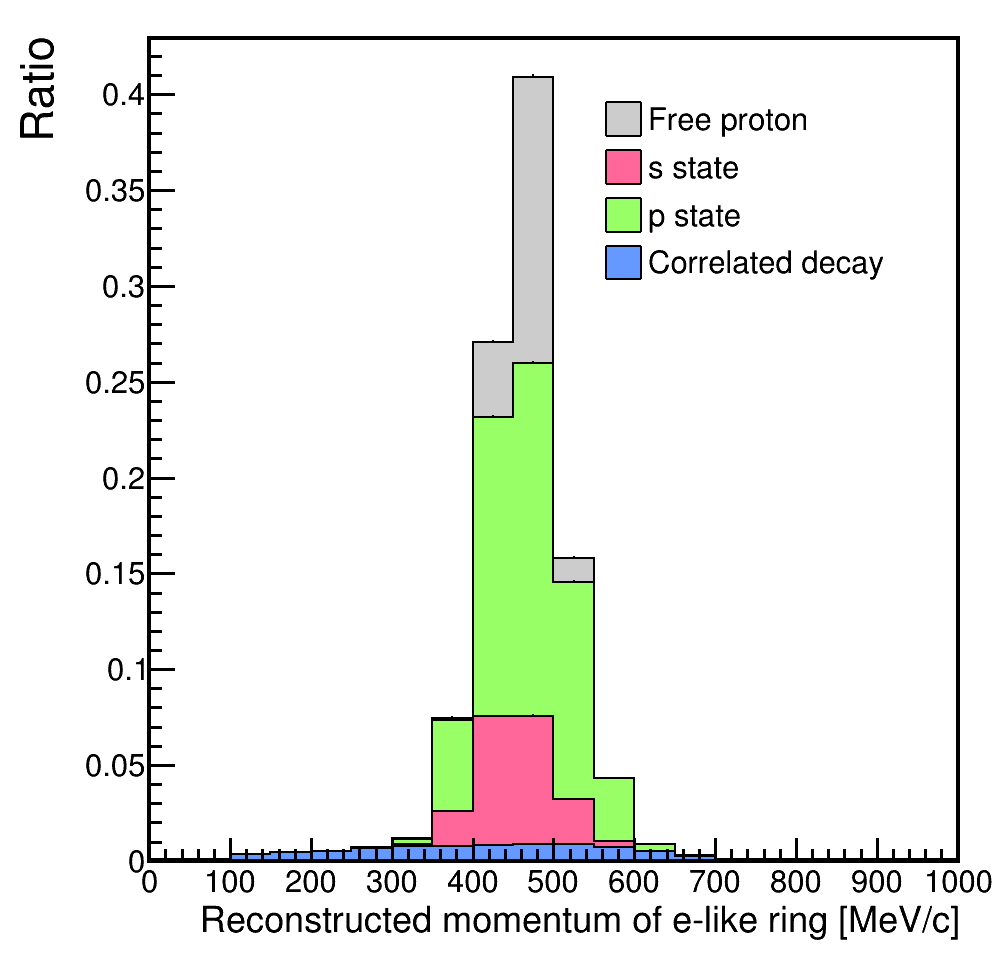}
    \includegraphics[width=.4\textwidth]{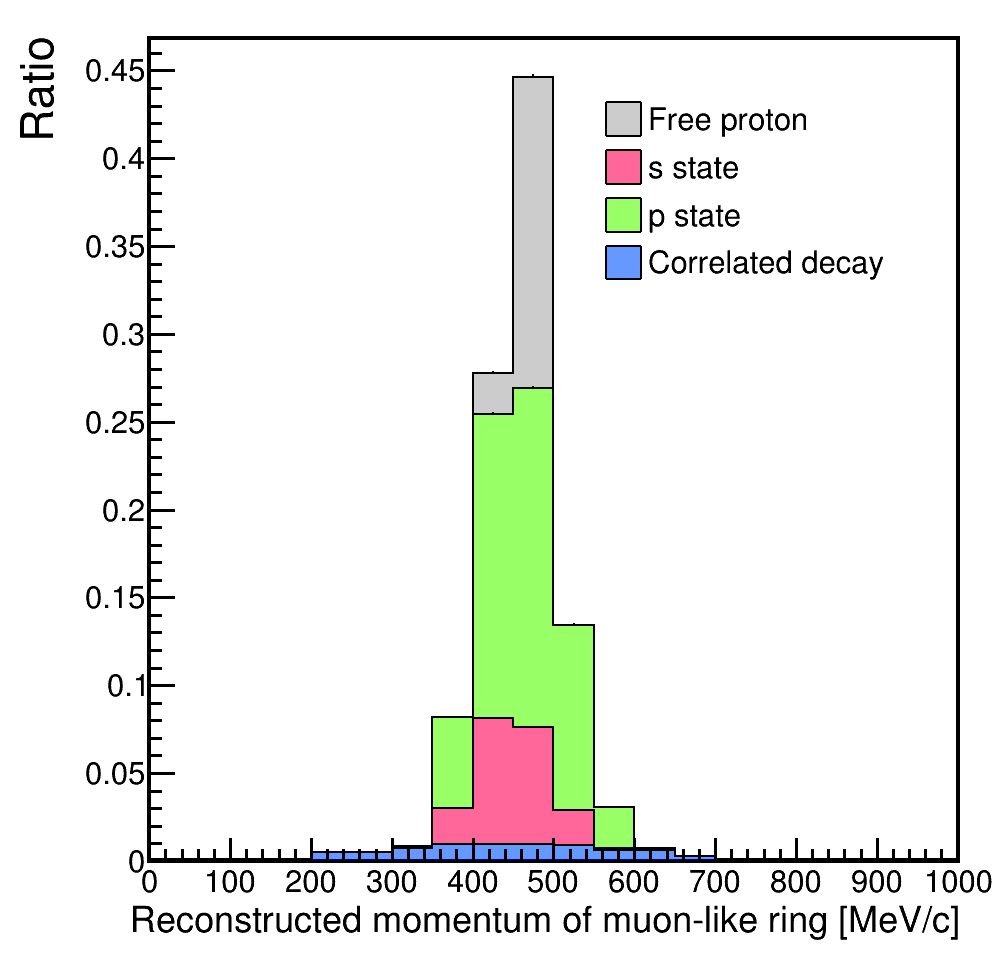}
    \caption{Reconstructed momentum of the $p\rightarrow e^{+}+X$ (top) and $p\rightarrow \mu^{+}+X$ (bottom) signals generated for SK-IV+SK-V after applying the selection criteria. The components of the free protons, bound protons ($s$ and $p$ states), and correlated decay are shown.} 
    \label{fig:signal}
\end{figure}

Note that the improved electronics used in the SK-IV and -V data periods allows for neutron identification via the 2.2~MeV $\gamma$ emitted when neutrons capture on protons (efficiency $\sim 20\%$). 
Though atmospheric neutrino interactions often produce neutrons and proton decay events do not, large uncertainties in the number of neutrons produced in atmospheric neutrinos limit the efficacy of a neutron selection criterion in the present analysis. 
Accordingly no such cut is included here. 
The reconstructed momentum of the proton decay signal samples generated for SK-IV+SK-V after applying the selection criteria are shown in Figure~\ref{fig:signal}. 

After applying the selection criteria, the signal detection efficiencies of the final signal samples for each decay mode are estimated.
The signal detection efficiency is defined as the fraction of events that pass the selection criteria compared with the total number of events generated with a true vertex position within the fiducial volume.  
Table~\ref{tab:table1} shows the signal detection efficiencies and atmospheric neutrino background rates of the decay modes, with each statistical uncertainty after applying each event selection step for SK-IV and -V. 
The collected SK-IV and -V data counts are also listed at each event selection step. 
Values from the previous search using SK-I to SK-III~\cite{PhysRevLett.115.121803, lXphd} are listed at the bottom as well. 
Compared to the efficiencies in SK-I to SK-III, the increase in efficiency observed in SK-IV and -V for $p\rightarrow \mu^{+}+X$ comes from the upgrade of the detector electronics introduced in Section II. 
The upgrade provides a 20$\%$ improvement in the detection of Michel electrons in this study~\cite{lXphd}.

\begin{table*}[]
\caption{\label{tab:table1}%
Signal detection efficiencies, atmospheric neutrino background rates and data from the $p\rightarrow e^{+}+X$ and $p\rightarrow \mu^{+}+X$ analyses after applying each event selection step in SK-IV and -V. 
Values from the previous search using SK-I to SK-III data are listed as well~\cite{PhysRevLett.115.121803, lXphd}. 
Errors here are statistical uncertainties only. 
Note that the atmospheric-$\nu$ (ATM-$\nu$) background represents the nominal MC number, prior to spectral fitting.}
\begin{tabular}{llccccccc}
\hline \hline
\multirow{3}{*}{Selection} &  & \multicolumn{3}{c}{$p\rightarrow e^{+}+X$}                                          &                      & \multicolumn{3}{c}{$p\rightarrow \mu^{+}+X$}                                           \\ \cline{3-5} \cline{7-9} 
                           &  & Signal               & ATM-$\nu$ rate       & Data                 &                      & Signal               & ATM-$\nu$ rate       & Data                 \\
                           &  & Efficiency ($\%$)          & livetime normalized &                      &                      & Efficiency ($\%$)          & livetime normalized &                      \\ \hline
SK-IV     &  & \multicolumn{1}{l}{} & \multicolumn{1}{l}{} & \multicolumn{1}{l}{} & \multicolumn{1}{l}{} & \multicolumn{1}{l}{} & \multicolumn{1}{l}{} & \multicolumn{1}{l}{} \\
FCFV pre-selection         &  & 98.1$\pm$0.1         & 24973$\pm$33            & 26542                &                      & 98.7$\pm$0.1         & 24973$\pm$33            & 26542                 \\
FCFV (i)                   &  & 98.1$\pm$0.1         & 19000$\pm$72             & 20154                &                      & 98.7$\pm$0.1         & 19000$\pm$72             & 20154                 \\
FCFV (i)-(ii)              &  & 96.3$\pm$0.1         & 14377$\pm$80             & 15096                &                      & 98.1$\pm$0.1         & 14377$\pm$80             & 15096                 \\
FCFV (i)-(iii)             &  & 95.7$\pm$0.1         & 7916$\pm$74             & 7923                 &                      & 97.3$\pm$0.1         & 6461$\pm$70             & 7173                 \\
FCFV (i)-(iv)              &  & 95.7$\pm$0.1         & 7184$\pm$72             & 7061                 &                      & 95.6$\pm$0.1         & 5038$\pm$64             & 5582                 \\
FCFV (i)-(v)               &  & 95.5$\pm$0.1         & 5791$\pm$67             & 5973                 &                      & 95.6$\pm$0.1         & 4545$\pm$61             & 5048                  \\ \hline
SK-V   &  & \multicolumn{1}{l}{} & \multicolumn{1}{l}{} & \multicolumn{1}{l}{} & \multicolumn{1}{l}{} & \multicolumn{1}{l}{} & \multicolumn{1}{l}{} & \multicolumn{1}{l}{} \\
FCFV pre-selection         &  & 97.5$\pm$0.1         & 3569$\pm$13            & 3861                & \multicolumn{1}{l}{} & 98.3$\pm$0.1         & 3569$\pm$13            & 3861                 \\
FCFV (i)                   &  & 97.5$\pm$0.1         & 2724$\pm$27             & 2950                & \multicolumn{1}{l}{} & 98.3$\pm$0.1         & 2724$\pm$27             & 2950                 \\
FCFV (i)-(ii)              &  & 95.8$\pm$0.1         & 2071$\pm$30             & 2194                & \multicolumn{1}{l}{} & 97.7$\pm$0.1         & 2071$\pm$30             & 2194                 \\
FCFV (i)-(iii)             &  & 95.2$\pm$0.1         & 1148$\pm$28             & 1168                 & \multicolumn{1}{l}{} & 96.9$\pm$0.1         & 923$\pm$26             & 1026                 \\
FCFV (i)-(iv)              &  & 95.2$\pm$0.1         & 1042$\pm$27             & 1034                 & \multicolumn{1}{l}{} & 95.2$\pm$0.1         & 719$\pm$24             & 806                  \\
FCFV (i)-(v)               &  & 94.9$\pm$0.1         & 840$\pm$26             & 862                 & \multicolumn{1}{l}{} & 95.2$\pm$0.1         & 650$\pm$23             & 727                  \\ \hline
SK-I FCFV (i)-(v)                   &  & 92.7$\pm$0.4         & 2810$\pm$47             & 2912                & \multicolumn{1}{l}{} & 77.2$\pm$0.7         & 1787$\pm$39             & 1821                 \\
SK-II FCFV (i)-(v)                   &  & 95.1$\pm$0.3         & 1483$\pm$34             & 1534                & \multicolumn{1}{l}{} & 77.7$\pm$0.7         & 940$\pm$28             & 918                 \\
SK-III FCFV (i)-(v)                   &  & 94.4$\pm$0.4         & 958$\pm$27             & 1051                & \multicolumn{1}{l}{} & 80.3$\pm$0.6         & 621$\pm$23             & 658                 \\ \hline \hline
\end{tabular}
\end{table*}

\begin{figure}[!htbp]
    \centering
    \includegraphics[width=.4\textwidth]{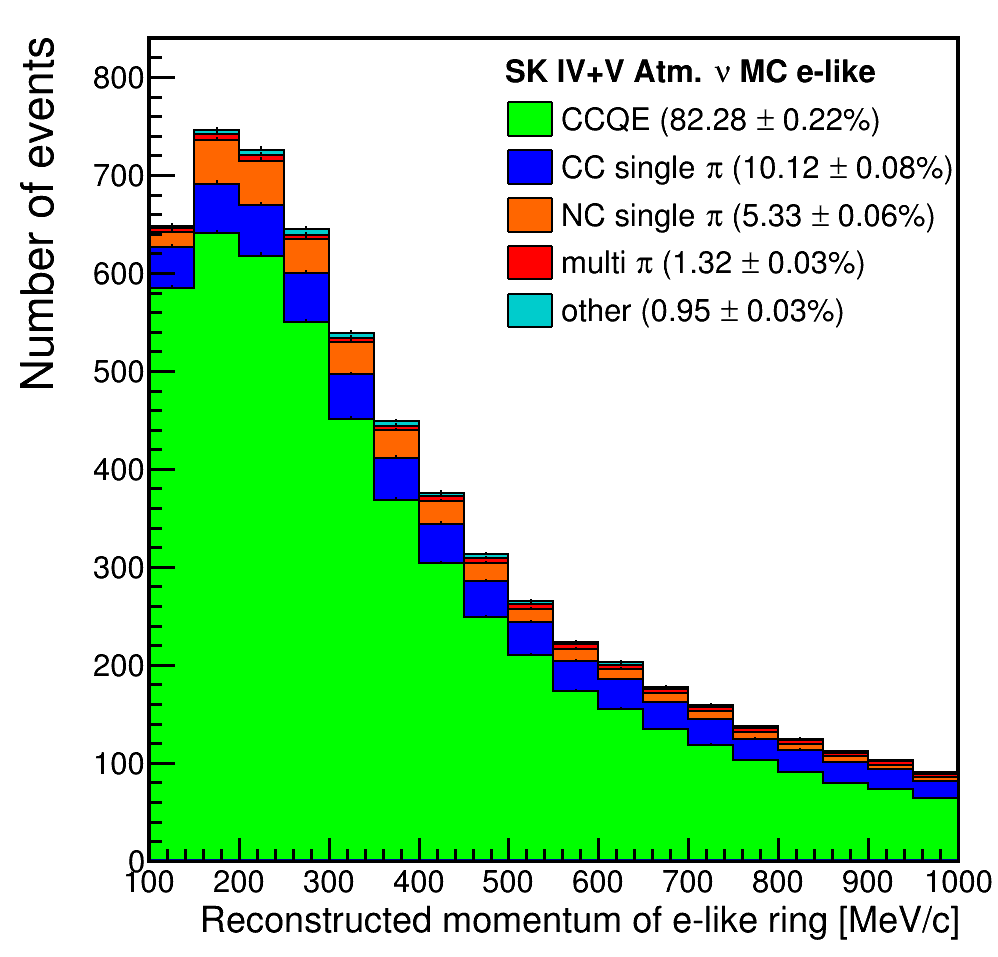}
    \includegraphics[width=.4\textwidth]{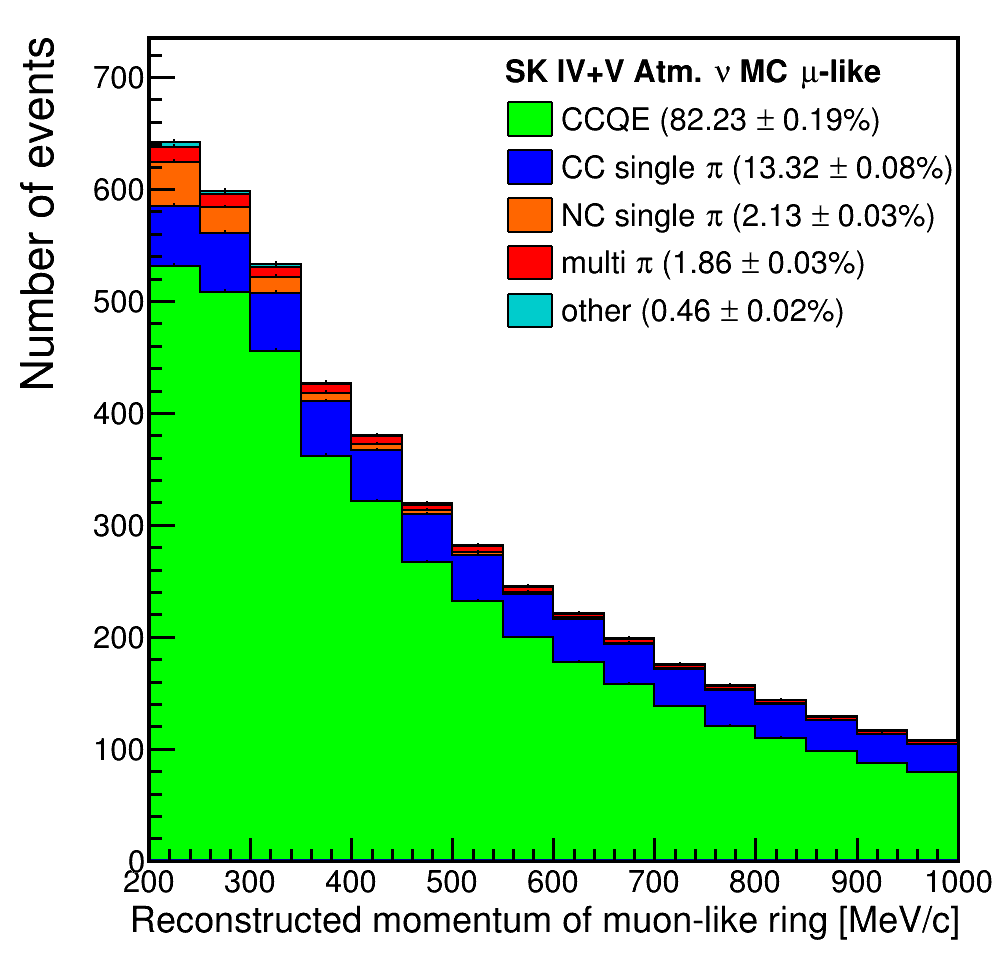}
    \caption{Neutrino interaction modes of the remaining reconstructed electron (top) and muon (bottom) momentum distributions of atmospheric neutrino backgrounds after applying all the selection criteria for SK-IV+SK-V. NC represents the neutral-current interaction.} 
    \label{fig:breakdown}
\end{figure}

The neutrino interaction modes contributing to the atmospheric neutrino backgrounds that survived the selection criteria are shown in Figure~\ref{fig:breakdown}. 
The primary background
source in both $p\rightarrow e^{+}+X$ and $p\rightarrow \mu^{+}+X$ is charged-current quasi-elastic (CCQE) neutrino interactions, comprising more than 80$\%$ of the remaining background events. 

After the event selection, a spectral fit is performed on the reconstructed charged lepton momentum distribution of the events to search for evidence of a signal. 
The fit is based on the $\chi^2$ below with systematic uncertainties included as terms modifying the event rate in each bin but subject to quadratic penalties ($``$pull terms$"$)~\cite{PhysRevD.66.053010}: 
\begin{align}
&\chi^{2}=2\sum_{i=1}^{nbins}\left(N^{exp}_{i}+N^{obs}_{i}\left[\ln\frac{N^{obs}_{i}}{N^{exp}_{i}}-1 \right]\right)+\sum_{j=1}^{N_{sys}}\left(\frac{\epsilon_{j}}{\sigma_{j}}\right)^{2}
\label{eq:chi1}\\
&N^{exp}_{i}=\left[N_{i}^{bkg}+\beta N_{i}^{sig}\right]\left(1+\sum_{j=1}^{N_{sys}}f_{i}^{j}\frac{\epsilon_{j}}{\sigma_{j}} \right).
\label{eq:chi2}
\end{align}
\noindent Here $i$ label bins of the reconstructed momentum distribution, $N^{obs}_{i}$, $N_{i}^{bkg}$, $N_{i}^{sig}$, and $N^{exp}_{i}$ are the numbers of observed data, background MC, signal MC, and total (signal$+$background) MC events, respectively. 
There are 18 bins for each SK period in the $p\rightarrow e^{+}+X$ analysis and 16 each for $p\rightarrow \mu^{+}+X$.
Both analyses use 50~MeV-wide bins and the signal and MC are binned separately.
The parameter $\beta$ is the signal normalization factor that is adjusted freely during the fit. 
When $\beta$ is equal to 1 approximately 220 signal events in total for each decay mode. 
The label $j$ indexes systematic errors and $f_{i}^{j}$ is the fractional change in the $i^{th}$ bin for each systematic error. 
The 1$\sigma$ uncertainty for error $j$ is represented by $\sigma_{j}$ and $\epsilon_{j}$ is the corresponding error parameter.
At each tested value of $\beta$ the $\epsilon_{j}$ are fitted to minimize $\chi^{2}$ according to $\partial \chi^{2}/\partial \epsilon_{j}=0$. 

\section{Systematic uncertainties}
The impact of systematic errors on this analysis is considered within the spectral fit through the $f_{i}^{j}$ coefficients described above. 
For simplicity errors are assumed to linearly affect the nominal bin content, $N_{i}^{0}$, such that the $f_{i}^{j}$s are defined as the fractional change in bin content under $\pm1\sigma_j$ variations in the systematic: 
\begin{equation}
f_{i}^{j}=\frac{N_{i}^{+\sigma_{j}}-N_{i}^{-\sigma_{j}}}{2N_{i}^{0}}.
\label{eq:fig}
\end{equation}
Since the signal and background bins are separated prior to fitting, this ensures that signal and background systematics are applied only to the signal and background bins, respectively. 

The systematic uncertainties can be divided into three categories: uncertainties relevant to the signal ($S$), uncertainties relevant to both the signal and background ($SB$), and uncertainties relevant to the background ($B$). 
The Fermi motion and nucleon--nucleon correlated decay uncertainties belong to signal-relevant systematics. 
The uncertainty on the Fermi motion is taken as the ratio of events passing the event selection assuming a spectral function model~\cite{nakamura_hiramatsu_kamae_muramatsu_izutsu_watase_1976} relative to the default Fermi gas model~\cite{PhysRevC.83.045501}.
On the other hand,  the uncertainty in the probability of a correlated decay is assumed to be 100$\%$, the same value used in the previous analysis~\cite{PhysRevLett.115.121803}. 
As shown in Figure~\ref{fig:signal} the fraction of correlated decay events passing the selection is small, indicating these errors are sub-dominant in the present analysis.

Uncertainties in the detector response and event reconstruction impact both signal and background events. 
This analysis adopts uncertainties from the SK atmospheric neutrino oscillation analysis~\cite{PhysRevD.109.072014} that impact the sub-GeV fully-contained samples from each data period (summarized in Table~\ref{tab:table4}).
Among these, the energy scale uncertainties, which shift the reconstructed momentum distributions directly, result in a change in the signal and background rate of less than 1\%. 
Other uncertainties in the detector response produced similar or smaller changes, but were included in the analysis for completeness.

The primary systematic sources affecting only background events stem from uncertainties the atmospheric neutrino flux, the interaction modeling, and the assumed neutrino oscillation parameters. 
Flux-related errors are evaluated on the basis of the uncertainties in hadronic interactions, air density profile~\cite{PhysRevD.75.043006}, and from model comparisons~\cite{PhysRevD.70.023006, BATTISTONI2003269, BATTISTONI2003291}. 
CCQE-related uncertainties are evaluated via a comparison of Fermi gas models~\cite{SMITH1972605,PhysRevC.83.045501}. 
Since CCQE events are the dominant background source in both decay modes, the systematic error on the CCQE cross-section shape has the largest effect on the sensitivity.
For single-meson production, uncertainties from the Rein--Sehgal model~\cite{REIN198179} and  from comparison with the Hernandez model~\cite{PhysRevD.76.033005} are used.  
Uncertainties in the oscillation of atmospheric neutrinos are taken from SK's measurements~\cite{PhysRevD.109.072014}.
Reference~\cite{sysdetail} provides more details on the systematic error model.
All systematic uncertainties considered in this analysis, together with their values at the best fit, are summarized in Table~\ref{tab:table3} for $S$, Table~\ref{tab:table4} for $SB$, and Table~\ref{tab:table5} for $B$.

\section{Results}
\begin{figure}[!htbp]
    \centering
    \includegraphics[width=.4\textwidth]{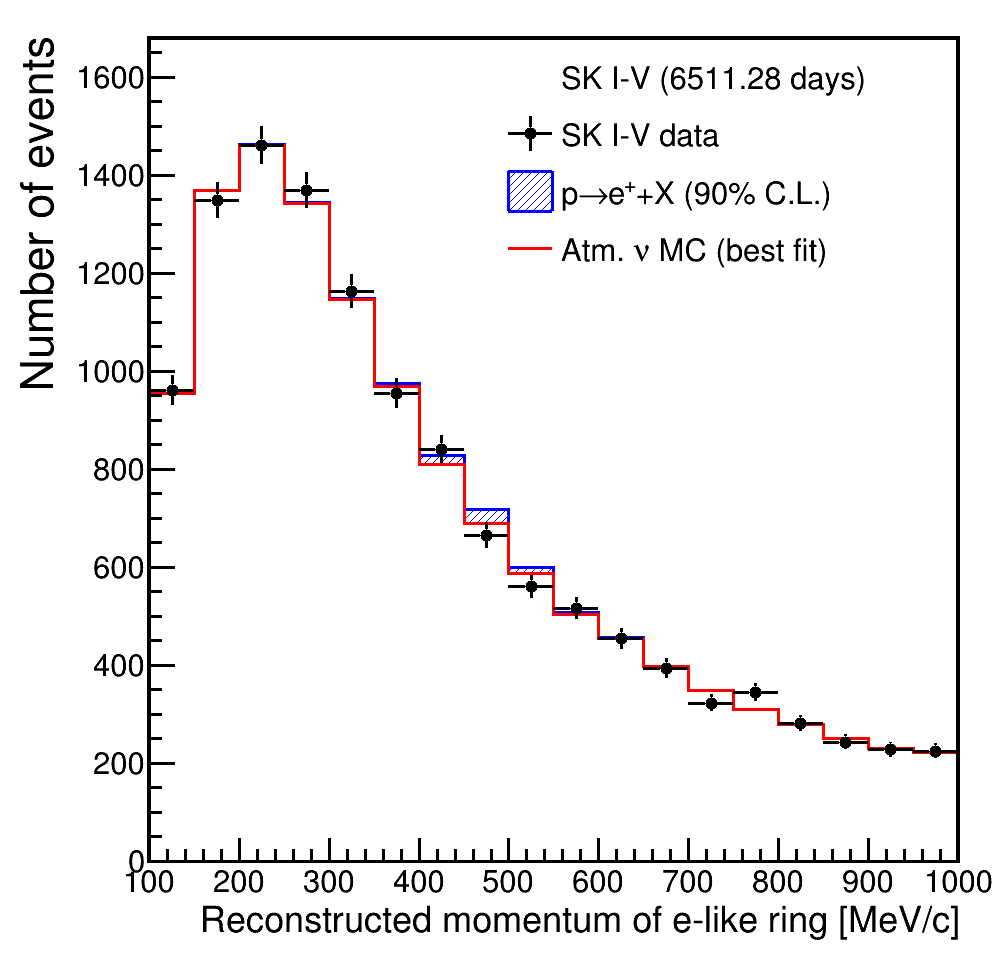}
    \includegraphics[width=.4\textwidth]{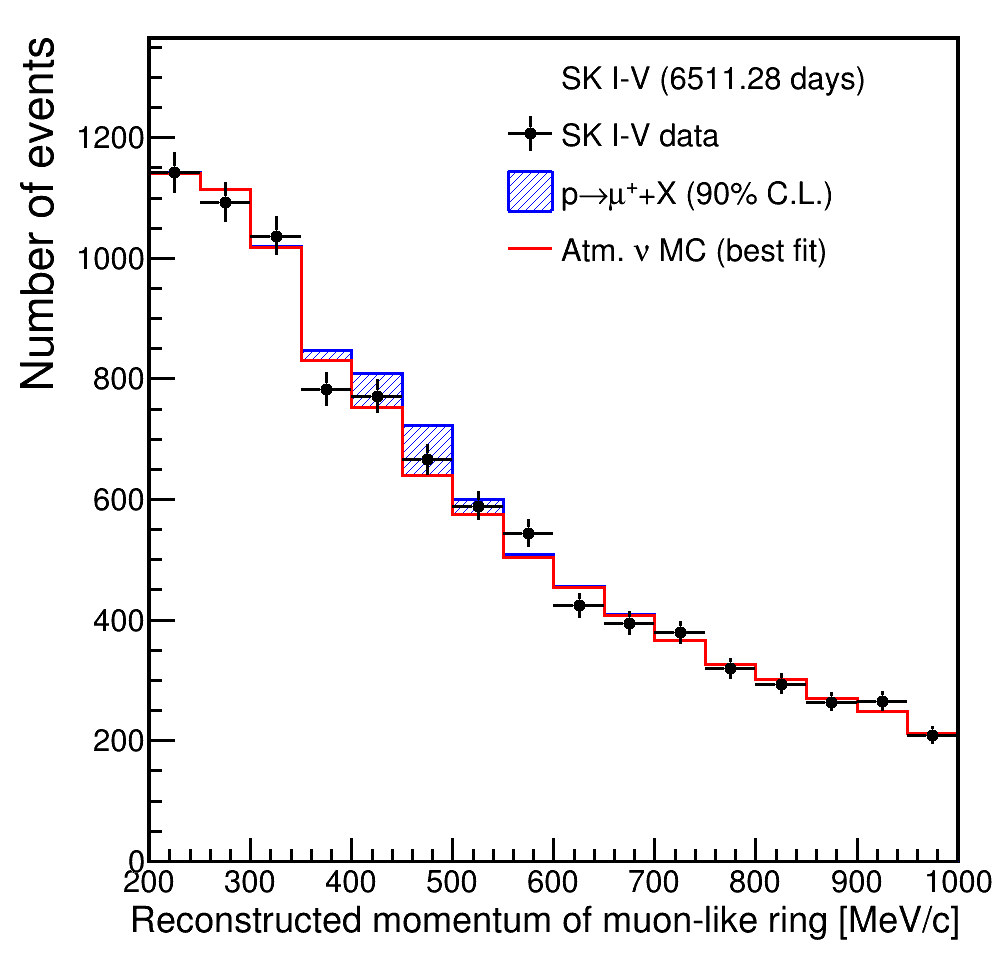}
    \caption{Reconstructed electron (top) and muon (bottom) momentum distributions of the SK data collected throughout the pure water phase (black dots), the best-fit distribution of the atmospheric neutrino background MC simulations (red line), and the allowed proton decay events (hatched histograms) fitting from the signal and background MC to the data at the 90$\%$ C.L.} 
    \label{fig:result_mom}
\end{figure}

The spectral fit minimized the $\chi^2$ above while adjusting systematic error parameters to achieve the best agreement between data and MC at each tested value of $\beta$. 
Following the fit, the value of $\beta$ with the smallest $\chi^2$ is deemed the best fit. 
The total number of predicted events at any $\beta$ is given by $\beta\times N_{signal}$, where $N_{signal}$ represents the total number of signal events described in Section V.
In the $p\rightarrow e^{+}+X$ there are 0 signal events at the best fit, whereas there are 91 fitted signal events for $p\rightarrow \mu^{+}+X$.
The latter does not exceed 3$\sigma$ significance and therefore both modes are consistent the no-proton-decay hypothesis. 
The number of allowed signal events at the 90$\%$ C.L., $N_{90}$, for $p\rightarrow e^{+}+X$ and $p\rightarrow \mu^{+}+X$ is 73 and 190 events, respectively. 
The blue-hatched histograms in Figure~\ref{fig:result_mom} show the distributions of the allowed signal events for each decay mode at the 90$\%$ C.L., respectively.
The $\chi^{2}$/dof of the best-fit for $p\rightarrow e^{+}+X$ and $p\rightarrow \mu^{+}+X$ are 87.88/88 and 104.00/78, respectively.
Note that the latter shows a larger value due to data-MC discrepancies (within 2$\sigma$) in the 350-400 and 550-600 MeV/c bins.  

Lower lifetime limits on the proton lifetime are calculated at the 90$\%$ C.L. as follows:
\begin{equation}
\begin{aligned}
\tau_{90}/\mathcal{B}=\frac{\sum_{i}\lambda_{i}\times\epsilon_{i}\times N_{proton}}{N_{90}},
\label{eq:lifetime}
\end{aligned}
\end{equation}
where $\mathcal{B}$ indicates the branching ratio of the proton decay mode, the label $i$ represents each SK period, $\lambda_{i}$ is the detector exposure (kton$\cdot$year), $\epsilon_{i}$ is the signal efficiency, $N_{proton}$ is the number of protons in the SK tank ($3.34\times10^{32}$ kton\textsuperscript{-1}), and $N_{90}$ is the number of allowed signal events at the 90$\%$ C.L..  
$N_{90}$ is computed from the 90$\%$ C.L. limit on $\beta$, which is determined using the constant critical value for a $\chi^2$ with one degree of freedom.

This analysis places lifetime limits of $1.72\times10^{33}$ and $0.61\times10^{33}$ years for $p\rightarrow e^{+}+X$ and $p\rightarrow \mu^{+}+X$, respectively.
These limits improve our previous result by factors of 2 and 1.5 times~\cite{PhysRevLett.115.121803}.
The expected sensitivity is $1.2\times10^{33}$ years for both modes. 
This number is computed using $N_{90}$ from a fit to the background-only MC and results in 105 and 97 events, respectively.
Compared to the previous study, the sensitivity has improved by a factor of 1.5. 

\section{conclusion}
Proton decays with invisible particles, such as $p\rightarrow l^{+}+X$, provide a unique opportunity to broadly explore baryon number violation and light new physics beyond the SM.
To approach such scenarios, the present work searched for proton decay into a single charged antilepton and a massless invisible particle using 401 kton$\cdot$years of pure water data from Super-Kamiokande. 
No significant statistical evidence of data with respect to the background is observed. 
Accordingly, new lower bounds on the partial lifetime of proton are set at $1.72\times10^{33}$ and $0.61\times10^{33}$ years for $p\rightarrow e^{+}+X$ and $p\rightarrow \mu^{+}+X$ at the 90$\%$ C.L., respectively. 
These limits improve on previous results by factors of 2 and 1.5, respectively. 

\begin{acknowledgments}
We gratefully acknowledge the cooperation of the Kamioka Mining and Smelting Company. The Super-Kamiokande experiment has been built and operated from funding by the Japanese Ministry of Education, Culture, Sports, Science and Technology; the U.S. Department of Energy; and the U.S. National Science Foundation. Some of us have been supported by funds from the National Research Foundation of Korea (NRF-2009-0083526, NRF-2022R1A5A1030700, NRF-2022R1A3B1078756, RS-2025-00514948) funded by the Ministry of Science, Information and Communication Technology (ICT); the Institute for Basic Science (IBS-R016-Y2); and the Ministry of Education (2018R1D1A1B07049158, 2021R1I1A1A01042256, RS-2024-00442775); the Japan Society for the Promotion of Science; the National Natural Science Foundation of China (Grants No. 12375100 and 12521007); the Spanish Ministry of Science, Universities and Innovation (grant PID2021-124050NB-C31); the Natural Sciences and Engineering Research Council (NSERC) of Canada; the Scinet and Digital Research of Alliance Canada; the National Science Centre (UMO-2018/30/E/ST2/00441 and UMO-2022/46/E/ST2/00336) and the Ministry of  Science and Higher Education (2023/WK/04), Poland; the Science and Technology Facilities Council (STFC) and Grid for Particle Physics (GridPP), UK; the European Union’s Horizon 2020 Research and Innovation Programme H2020-MSCA-RISE-2018 JENNIFER2 grant agreement no.822070, H2020-MSCA-RISE-2019 SK2HK grant agreement no. 872549; and European Union's Next Generation EU/PRTR  grant CA3/RSUE2021-00559; the National Institute for Nuclear Physics (INFN), Italy.
\end{acknowledgments}

\begin{table*}[]
\caption{\label{tab:table4}%
Systematic uncertainties relevant to both signal and background categories. The 1$\sigma$ uncertainty values in percent and best-fit pull values for each decay mode corresponding to each detector dependence are listed.}
\begin{tabular}{lcccccccccccccccc}
\hline \hline
\multirow{2}{*}{$SB$ systematic}     &  & \multicolumn{2}{c}{SK-I}                     &  & \multicolumn{2}{c}{SK-II}                    &  & \multicolumn{2}{c}{SK-III}                   &  & \multicolumn{2}{c}{SK-IV}                    &  & \multicolumn{2}{c}{SK-V}                     &  \\ \cline{3-4} \cline{6-7} \cline{9-10} \cline{12-13} \cline{15-16}
                                     &  & 1$\sigma$ ($\%$) & Fit pull &  & 1$\sigma$ ($\%$) & Fit pull &  & 1$\sigma$ ($\%$) & Fit pull &  & 1$\sigma$ ($\%$) & Fit pull &  & 1$\sigma$ ($\%$) & Fit pull &  \\ \hline
$p\rightarrow e^{+}+X$               &  &                  &                           &  &                  &                           &  &                  &                           &  &                  &                           &  &                  &                           &  \\
FC reduction                         &  & 0.2              & $-$0.022                    &  & 0.2              & $-$0.014                    &  & 0.8              & 0.076                     &  & 1.3              & 0.084                     &  & 1.7              & 0.029                     &  \\
Fiducial volume                      &  & 2                & $-$0.225                    &  & 2                & $-$0.145                    &  & 2                & 0.190                     &  & 2                & 0.129                     &  & 2                & 0.034                     &  \\
Non-$\nu$ background ($e$-like)      &  & 1                & $-$0.010                    &  & 1                & $-$0.021                    &  & 1                & 0.039                     &  & 1                & 0.001                     &  & 1                & 0.016                     &  \\
Ring separation                      &  & 10               & $-$0.026                    &  & 10               & $-$0.074                    &  & 10               & $-$0.028                    &  & 10               & $-$0.007                    &  & 10               & 0.200                     &  \\
PID 1 ring                           &  & 1                & 0.056                     &  & 1                & $-$0.029                    &  & 1                & $-$0.019                    &  & 1                & $-$0.019                    &  & 1                & $-$0.004                    &  \\
Michel-e tagging                     &  & 10               & 0.013                     &  & 10               & 0.009                     &  & 10               & $-$0.012                    &  & 10               & $-$0.004                    &  & 10               & $-$0.022                    &  \\
Energy scale                   &  & 3.3              & 0.052                     &  & 2                & $-$1.067                    &  & 2.4              & 0.180                     &  & 2.1              & 0.659                     &  & 1.8              & 0.674                     &  \\
Up/Down energy calibration &  & 0.6              & $-$0.009                    &  & 1.1              & 0.030                     &  & 0.6              & $-$0.013                    &  & 0.5              & $-$0.020                    &  & 0.7              & 0.059                     &  \\ \hline
$p\rightarrow \mu^{+}+X$             &  &                  &                           &  &                  &                           &  &                  &                           &  &                  &                           &  &                  &                           &  \\
FC reduction                         &  & 0.2              & $-$0.073                    &  & 0.2              & $-$0.097                    &  & 0.8              & $-$0.153                    &  & 1.3              & 0.099                     &  & 1.7              & 0.094                     &  \\
Fiducial volume                      &  & 2                & $-$0.726                    &  & 2                & $-$0.972                    &  & 2                & $-$0.381                    &  & 2                & 0.152                     &  & 2                & 0.111                     &  \\
Non-$\nu$ background ($\mu$-like)    &  & 1                & $-$0.360                    &  & 1                & $-$0.165                    &  & 1                & $-$0.117                    &  & 1                & 0.031                     &  & 1                & 0.049                     &  \\
Ring separation                      &  & 10               & 0.244                     &  & 10               & $-$0.428                    &  & 10               & 0.222                     &  & 10               & 0.058                     &  & 10               & $-$0.122                    &  \\
PID 1 ring                           &  & 1                & $-$0.182                    &  & 1                & 0.194                     &  & 1                & $-$0.057                    &  & 1                & 0.030                     &  & 1                & 0.009                     &  \\
Michel-e tagging                     &  & 10               & $-$0.545                    &  & 10               & $-$0.729                    &  & 10               & $-$0.286                    &  & 10               & 0.975                     &  & 10               & 0.164                     &  \\
Energy scale                   &  & 3.3              & $-$0.308                    &  & 2                & $-$0.156                    &  & 2.4              & $-$0.306                    &  & 2.1              & $-$0.063                    &  & 1.8              & $-$0.217                    &  \\
Up/Down energy calibration &  & 0.6              & 0.147                     &  & 1.1              & 0.079                     &  & 0.6              & $-$0.061                    &  & 0.5              & $-$0.016                    &  & 0.7              & $-$0.072                    &  \\ \hline \hline
\end{tabular}
\end{table*}

\begin{table}[!htbp]
\caption{\label{tab:table3}%
Systematic uncertainties relevant to the signal category. The 1$\sigma$ uncertainty values in percent and best-fit pull values for each decay mode are listed. Note that since the signal yield at the best-fit point for $p\rightarrow e^{+}+X$ is exactly zero, its Fermi motion and correlated decay errors make no contribution.}
\centering
\begin{tabular}{lccccc}
\hline \hline
\multirow{2}{*}{$S$ systematic} &\multirow{2}{*}{1$\sigma$ ($\%$)} & &\multicolumn{3}{c}{Fit pull} \\
                                & & & $p\rightarrow e^{+}+X$ & & $p\rightarrow \mu^{+}+X$ \\ \hline
Fermi motion     & 10           & & 0.000 & & $-$0.033 \\ 
Correlated decay & 100          & & 0.000 & & $-$0.066 \\ \hline \hline
\end{tabular}
\end{table}

\begin{table*}[]
\caption{\label{tab:table5}%
Systematic uncertainties relevant to the background category. The 1$\sigma$ uncertainty values in percent and best-fit pull values for each decay mode are listed.}
\begin{tabular}{lllccccc}
\hline \hline
\multirow{2}{*}{$B$ systematic}                                                            &                              & \multicolumn{1}{l}{} & \multirow{2}{*}{1$\sigma$ ($\%$)} &  & \multicolumn{3}{c}{Fit pull}                         \\
                                                                                           &                              & \multicolumn{1}{l}{} &                                   &  & $p\rightarrow e^{+}+X$ &  & $p\rightarrow \mu^{+}+X$ \\ \hline
\multirow{2}{*}{Flux normalization} & $E_{\nu}>1$ GeV              &  & 25\footnote{Uncertainty is 7$\%$ up to 10 GeV, linearly increases with log$E_{\nu}$ from 7 to 12$\%$ at 10--100 GeV, and then to 20$\%$ at 1 TeV.}                                &  & 0.090                  &  & $-$0.229                   \\
& $E_{\nu}<1$ GeV              &  & 15\footnote{Uncertainty linearly decreases with log$E_{\nu}$ from 25 to 7$\%$ at 0.1--1 GeV.}                                &  & $-$0.071                 &  & $-$0.064                   \\
\multirow{3}{*}{$(\nu_{\mu}+\overline{\nu}_{\mu})/(\nu_{e}+\overline{\nu}_{e})$ ratio} & $E_{\nu}<1$ GeV              &                      & 2                                 &  & $-$0.002                 &  & 0.026                    \\
                                                                                           & $1<E_{\nu}<10$ GeV           &                      & 3                                 &  & 0.004                  &  & $-$0.007                   \\
                                                                                           & $E_{\nu}>10$ GeV             &                      & 5\footnote{Uncertainty linearly increases with log$E_{\nu}$ from 5 to 30$\%$ at 30 GeV--1 TeV.}                                 &  & $-$0.001                 &  & 0.000                    \\
\multirow{3}{*}{$\overline{\nu}_{e}/\nu_{e}$ ratio}                               & $E_{\nu}<1$ GeV              &                      & 5                                 &  & 0.074                  &  & 0.000                    \\
                                                                                           & $1<E_{\nu}<10$ GeV           &                      & 5                                 &  & $-$0.027                 &  & $-$0.003                   \\
                                                                                           & $E_{\nu}>10$ GeV             &                      & 8\footnote{Uncertainty linearly increases with log$E_{\nu}$ from 8 to 20$\%$ at 100 GeV--1 TeV.}                                 &  & 0.001                  &  & 0.000                    \\
\multirow{3}{*}{$\overline{\nu}_{\mu}/\nu_{\mu}$ ratio}                           & $E_{\nu}<1$ GeV              &                      & 2                                 &  & 0.000                  &  & 0.001                    \\
                                                                                           & $1<E_{\nu}<10$ GeV           &                      & 2                                 &  & $-$0.006                 &  & $-$0.024                   \\
                                                                                           & $E_{\nu}>10$ GeV             &                      & 6\footnote{Uncertainty linearly increases with log$E_{\nu}$ from 6 to 40$\%$ at 50 GeV--1 TeV.}                                 &  & 0.000                  &  & 0.000                    \\
Neutrino path length                                                                       &                              &                      & 10                                 &  & 0.001                  &  & $-$0.070                   \\
\multirow{5}{*}{Solar activity}                                                            & SK-I                         &                      & 20                                &  & 0.142                  &  & $-$0.075                   \\
                                                                                           & SK-II                        &                      & 50                                &  & 0.218                  &  & $-$0.657                   \\
                                                                                           & SK-III                       &                      & 20                                &  & 0.106                  &  & $-$0.032                   \\
                                                                                           & SK-IV                        &                      & 7                                 &  & $-$0.162                 &  & 0.120                    \\
                                                                                           & SK-V                         &                      & 20                                &  & 0.002                  &  & 0.069                    \\
Up/Down ratio                                                                              &                              &                      & 1                                 &  & 0.000                  &  & $-$0.044                   \\
Horizontal/Vertical ratio                                                                  &                              &                      & 1                                 &  & $-$0.004                 &  & $-$0.017                   \\
$K/\pi$ ratio                                                                              &                              &                      & 10\footnote{Uncertainty is 5$\%$ up to 100 GeV, linearly increases with log$E_{\nu}$ from 5$\%$ to 5$\%$ at 100 GeV--1 TeV.}                                 &  & $-$0.004                 &  & 0.010                    \\
\multirow{5}{*}{CCQE cross section}                                                                  & shape                              &                      & 10                                &  & 0.543                  &  & 0.541                    \\
                                                        & Sub-GeV                      &                      & 10                                &  & $-$0.041                 &  & $-$0.013                   \\
                                                                                           & Multi-GeV                    &                      & 10                                &  & 0.175                  &  & 0.422                    \\
                                                                                           & $\overline{\nu}/\nu$ &                      & 10                                &  & 0.117                  &  & 0.092                    \\
                                                                                           & $\nu_{\mu}/\nu_{e}$  &                      & 10                                &  & $-$0.315                 &  & 0.115                    \\
DIS model difference                                                                       &                              &                      & 10                                &  & $-$0.059                 &  & $-$0.153                   \\
DIS cross section                                                                          &                              &                      & 10                                &  & $-$0.010                 &  & $-$0.027                   \\
DIS hadron multiplicity                                                                    &                              &                      & 10                                &  & $-$0.006                 &  & $-$0.047                   \\
\multirow{4}{*}{DIS $Q^{2}$}                                                              & High W                       &                      & 10                                &  & 0.005                  &  & $-$0.013                   \\
                                                                                           & Low W                        &                      & 10                                &  & $-$0.004                 &  & $-$0.010                   \\
                                                                                           & Vector                       &                      & 10                                &  & 0.014                  &  & 0.029                    \\
                                                                                           & Axial                        &                      & 10                                &  & 0.025                  &  & $-$0.006                   \\
Coherent $\pi$ production cross section                                                    &                              &                      & 100                               &  & 0.038                  &  & 0.064                    \\
\multirow{5}{*}{Single meson production}                                                   & $\pi^{0}/\pi^{\pm}$        &                      & 40                                &  & 0.091                  &  & 0.291                    \\
                                                                                           & $\overline{\nu}/\nu$ &                      & 10                                &  & 0.419                  &  & 0.186                    \\
                                                                                           & Axial coupling               &                      & 10                                &  & $-$0.384                 &  & $-$0.186                   \\
                                                                                           & $C_{A}^{5}$                  &                      & 10                                &  & $-$0.260                 &  & $-$0.200                   \\
                                                                                           & Background                   &                      & 10                                &  & $-$0.122                 &  & 0.000                    \\
NC/CC ratio                                                                                &                              &                      & 20                                &  & $-$0.091                 &  & $-$0.222                   \\
Axial mass                                                                                 &                              &                      & 10                                &  & 0.419                  &  & 0.587                    \\
Meson exchange current on/off                                                              &                              &                      & 10                                &  & 0.243                  &  & 0.206                    \\
Matter effect                                                                              &                              &                      & 6.8                               &  & 0.001                  &  & $-$0.007                   \\
sin$^{2}(\theta_{13})$                                                                     &                              &                      & 0.07                              &  & $-$0.003                 &  & $-$0.006                   \\
sin$^{2}(\theta_{12})$                                                                     &                              &                      & 1.3                               &  & 0.001                  &  & $-$0.022                   \\
$\Delta$m$^{2}_{21}$                                                                        &                              &                      & 0.00018                           &  & $-$0.031                 &  & $-$0.023                   \\
FSI max set                                                                                &                              &                      & 10                                &  & $-$0.472                 &  & 0.924                    \\
FSI min set                                                                                &                              &                      & 10                                &  & 0.200                  &  & $-$0.176                   \\
Hadron simulation\footnote{Hadron simulation only exists for $p\rightarrow \mu^{+}+X$.}                                                                          &                              &                      & 10                                &  &                        &  & $-$0.111                   \\ \hline \hline
\end{tabular}
\end{table*}

\bibliography{apssamp}

\end{document}